\documentclass[12pt]{spieman}  % 12pt font required by SPIE;
\usepackage{tocloft}
\usepackage{amsmath,amsfonts,amssymb}
\usepackage{mathabx}
\usepackage{deluxetable}
\usepackage{graphicx}
\usepackage{setspace}
\usepackage{float}
\usepackage{hhline}
\usepackage{lineno}
\newcommand{\rom}[1]{\uppercase\expandafter{\romannumeral #1\relax}}

\def\gappeq{\mathrel{ \rlap{\raise.5ex\hbox{$>$}}
                      {\lower.5ex\hbox{$\sim$}}  } }

\title{Optimized Bandpasses for the Habitable Worlds Observatory's ExoEarth Survey}

\author[a]{Christopher C. Stark}
\author[a,b]{Natasha Latouf}
\author[a]{Avi M. Mandell}
\author[a]{Amber Young}

\affil[a]{NASA Goddard Space Flight Center, Greenbelt, MD 20771, USA}
\affil[b]{George Mason University, 4400 University Dr, Fairfax, VA 22030, USA}

\cftpagenumbersoff{figure}
\cftpagenumbersoff{table} 
\begin{document} 
\maketitle

\begin{abstract}

A primary scientific goal of the future Habitable Worlds Observatory will be the direct detection and characterization of Earth-like planets. Estimates of the exoplanet yields for this concept will help guide mission design through detailed trade studies. It is therefore critical that yield estimation codes optimally adapt observations to the mission's performance parameters to ensure accurate trade studies. To aid in this, we implement wavelength optimization in yield calculations for the first time, allowing the yield code to determine the ideal detection and characterization bandpasses. We use this new capability to confirm the observational wavelength assumptions made for the LUVOIR-B study, namely that the optimum detection wavelength is 500 nm for the majority of targets and the optimum wavelength to detect water is near 1000 nm, given LUVOIR-B's assumed instrument performance parameters. We show that including the wavelength dependent albedo of an Earth twin as a prior provides no significant benefit to the yields of exoEarth candidates and caution against tuning observations to modern Earth twins. We also show that coronagraphs whose inner working angles are similar to step functions may benefit from wavelength optimization and demonstrate how wavelength-dependent instrument performance can impact the optimum wavelengths for detection and characterization. The optimization methods we implement automate wavelength selection and remove uncertainties regarding these choices, helping to adapt the observations to the instrument's performance parameters.

\end{abstract}

% Include a list of up to six keywords after the abstract

\begin{spacing}{1}

\section{Introduction}
\label{intro}

The Astro2020 Decadal Survey identified the detection and characterization of 25 potentially Earth-like planets as a primary science driver for NASA's next flagship mission, recommending that ``NASA should embark on a program to realize a mission to search for biosignatures from a robust number of about $\sim$25 habitable zone planets and to be a transformative facility for general astrophysics"\cite{astro2020}. The design of this mission, now known as the Habitable Worlds Observatory (HWO), will therefore be guided in part by estimates of the yield of potentially Earth-like planets. While it is important for the absolute yields from these calculations to be as precise as possible (e.g., by retiring astrophysical uncertainties), trade studies will require accurate \emph{relative} yields. To do this, yield calculations should strive to optimally adapt their simulated observations to the performance parameters of a given mission concept. As an example that will be explored throughout this paper, prescribing 950 nm as the required wavelength for HWO's water vapor search (as opposed to shorter wavelengths) may not be an ideal choice if the end-to-end throughput is low near 1 $\mu$m.

Ref.~\citenum{stark2014} showed the benefit of letting the yield code make optimized decisions. Doing so allows missions to take advantage of astronomical and observational degrees of freedom (e.g., the diameter and luminosity of the target stars, the depth of search of each star, etc.) to ensure that the observations adapt to the mission being studied. This resulted in improved maximization of yields and well-informed yield sensitivities, including a surprisingly weak sensitivity to exozodiacal dust\cite{stark2014}. Observation optimization is a crucial aspect of yield calculations if our goal is to optimize the design of HWO.  

Exoplanet yield calculations for high contrast imaging missions have been in development for roughly two decades, initially created to study the Terrestrial Planet Finder mission concept\cite{brown2004,levine2009}. Fundamentally, all exoplanet yield calculations work by adopting a high level description of a mission and simulating the observation of planetary systems assuming a fixed amount of mission time. After introducing the concept of obscurational completeness\cite{brown2004}, Ref.~\citenum{brown2005} immediately recognized that developing the numerical machinery to simulate the instrument performance was only half of the problem---optimizing the observations was key to maximizing the completeness of a mission. Ref.~\citenum{brown2005} was the first to optimize observations by focusing on which stars were selected. Stars were chosen based on prioritizing a benefit:cost metric, where the benefit was the completeness and the cost was the exposure time.

In addition to selecting targets, exposure times also had to be chosen. Ref.~\citenum{brown2005} initially explored optimizing exposure times by varying the search depth of the observation, $\Delta {\rm mag_{obs}}$, though this parameter was kept constant for all stars. Ref.~\citenum{turnbull2012} later improved on this by optimizing the search depth on a star-by-star basis, taking advantage of the fact that an Earth twin's differential magnitude compared to its host star, $\Delta {\rm mag}$, varies linearly with $\log{\left(L_{\star}/L_{\odot}\right)}$ due to the change in the habitable zone (HZ) separation, where $L_{\star}$ and $L_{\odot}$ are the bolometric luminosities of the star and Sun, respectively. Ref.~\citenum{hunyadi2007} determined the ideal exposure time optimization method by noting that the slope of the completeness vs time curve should be equal for all observations under the assumption of maximized productivity, though this was only implemented for a single visit to each star. Ref.~\citenum{stark2015} then expanded this concept to multiple visits to each star, self-consistently optimizing the exposure times on an observation-by-observation basis, as well as the number of and delay time between visits to maximize completeness.

Additional optimizations still existed. The Segmented Coronagraph Design and Analysis Study\footnote{\url{https://exoplanets.nasa.gov/exep/technology/SCDA/}} used yield calculations to guide instrument design. It was quickly realized that multiple coronagraphs could work together and suites of coronagraphs were designed to maximize yield\cite{stlaurent2018}. Ref.~\citenum{stark2019} added coronagraph selection optimization to yield calculations on a star-by-star basis, allowing the code to select the optimum coronagraph for each star, further maximizing completeness.

Recently yield studies have focused on optimization based on detailed orbital information \cite{brown2015,spohn2022}. Ref.~\citenum{morgan2021a} showed that prior knowledge of the orbits of potentially Earth-like planets from precursor radial velocity observations can increase the speed at which a mission can spectrally characterize exoEarths. Ref.~\citenum{morgan2021b} considered using orbit estimates of detected planets to adjust the cadence of observations for a given star.

%There are many additional aspects of observations yet to be optimized. Most of these require increasing levels of fidelity, including the PSF subtraction method for high contrast imaging and the point source detection algorithm. While yield calculations may benefit from and help guide some of these choices, much development and simulation work needs to be done before these details can be incorporated into yield studies with any degree of confidence.

Here we return to basics to address a fundamental aspect of observation that has so far gone un-optimized: wavelength. There are many reasons to expect that changing the detection wavelength should have an impact on yields. First, the inner working angle (IWA) of a coronagraph scales as $\lambda/D$, where $\lambda$ is wavelength and $D$ is telescope primary mirror diameter. By moving to shorter wavelengths, HZs around more distant stars and later type stars will become accessible. Second, the solid angle of the PSF is proportional to $\lambda^2$. The contribution of background noise scales with the solid angle of the PSF, so shorter wavelengths should decrease all astrophysical noise terms.

On the other hand, the stellar spectrum may push us to longer wavelengths. It is often assumed that we should image Earth-like planets near V band because that's where Sun-like stars are ``brightest." However, what the exposure time equation actually cares about is the number of photons collected, which tends to peak at longer wavelengths. In addition, Rayleigh scattering causes the Earth's reflectance to change significantly with wavelength, peaking near $\sim$350 nm. On top of all of this, the system end-to-end throughput varies with wavelength; the throughput for the LUVOIR mission concept peaks near $\sim$500 nm \cite{stark2019}. 

There are also wavelength choices when it comes to the spectral characterization of a potentially habitable planet. The LUVOIR and HabEx mission concept studies both adopted 1000 nm as the effective characterization wavelength, motivated by the desire to observe the deep water vapor absorption band at 950 nm\cite{luvoirfinalreport,habexfinalreport}. However, there are additional water bands at shorter wavelengths. Although these absorption features are not as prominent, they may be favorable for stars with compact HZs.

In reality, our choice of observation wavelengths may be determined by our desire to measure the planet's color, or sample its continuum over a broad range of wavelengths. However, in this study we pose a simple question: if our desire is to maximize the yield of potentially Earth-like planets, what is the ideal wavelength for our observations?

In this paper we address wavelength optimization on a star-by-star basis in yield calculations for the first time. In Section \ref{wavelength_opt_section} we describe how we updated the Altruistic Yield Optimization (AYO) code to include wavelength optimization for both detection and characterization observations. Then, in Section \ref{results_section} we incorporate wavelength optimization in increasing levels of fidelity to illustrate the impact of each of the factors described above. Finally, we investigate additional scenarios in which wavelength optimization is necessary and offer comments on the implications of wavelength optimization on mission design.

The methods we develop for this study can be used to evaluate the utility of different bandpasses and inform operations of a mission. For the former, we can adopt wavelength-independent mission performance parameters and ask the yield code to determine the most productive bandpass to achieve a particular science goal. This is possible as long as we know the observational requirements to achieve our goal as a function of wavelength, which we provide an example of in Section \ref{section:full_opt} for the detection of water vapor. For the latter, we can adopt realistic wavelength-dependent performance parameters describing a particular set of technologies, and ask the yield code to determine the ideal bandpass for observations on a star-by-star basis, which we provide an example of in Section \ref{section:utility}.

\section{Numerical wavelength optimization\label{wavelength_opt_section}}

The Altruistic Yield Optimization algorithm works by distributing $\sim10^5$ planets around each star, calculating their exposure times, sorting them to determine completeness as a function of time, and then optimizing the exposure time and epoch of every observation to maximize yield\cite{stark2014,stark2015}. Ideally, wavelength would be optimized simultaneously with exposure time and epoch optimization, such that wavelength was selected on an observation-by-observation basis. However, this leads to a number of numerical challenges. First, updating exposure times while performing optimization would slow down the optimization dramatically; the AYO code's efficiency comes from calculating all exposure times once, then performing optimization after the fact. One could calculate exposure times for all planets at all wavelengths ahead of time and keep them accessible for later optimization, but this would lead to a memory bottleneck. Assuming a 4-byte float for exposure time, calculated for $\sim$2k stars and up to four coronagraph designs, each wavelength slice would require $\sim$4 GB of RAM, effectively limiting the code to run only on high end clusters. Even if one were to do this, the optimization would slow down by a factor of $20$ assuming only five wavelength slices and four coronagraph designs. This run time increase is not consistent with our goals for the code.

To solve this issue, we optimize wavelength selection on a star-by-star basis instead of an observation-by-observation basis. We expect this to result in satisfactory optimization for the vast majority of stars. This assumption may break down for scenarios in which the HZ width exceeds the width of the coronagraph field of view (FOV). In these cases one could increase completeness by piecing together multiple observations at different wavelengths. For example, observing at a longer wavelength would allow us to cover the HZ planets at quadrature, and observing at a shorter wavelength would provide access to gibbous phase planets that were behind the coronagraph IWA at the longer wavelength. As no coronagraph designed for the LUVOIR or HabEx mission concepts had a FOV narrower than the HZ\cite{luvoirfinalreport,habexfinalreport}, we expect the numerical challenges to greatly outweigh the practical benefit from enabling this.

To employ wavelength optimization on a star-by-star basis, we adopt a vector of discrete wavelength options and apply the same method established for coronagraph optimization by Ref.~\citenum{stark2019}. For each star, we calculate exposure times for the $10^5$ synthetic planets for a given combination of coronagraph design and detection wavelength. Once calculated, we sort all planets by exposure time to calculate first-visit completeness, $C$, as a function of exposure time, $\tau$. We then calculate the benefit:cost metric for the star, $C/\tau$, and determine its peak value. We repeat this process for each combination of coronagraph design and detection wavelength, updating all relevant parameters of the exposure time equation as we go. In the end we select the combination of coronagraph design and wavelength with the highest peak value of first-visit $C/\tau$. 

This method frees us from needing access to exposure times for all coronagraphs and wavelengths simultaneously, greatly relaxing memory constraints. Assuming no numerical overheads, the method described above would increase the run time of the exposure time calculator by a factor of $N$, where $N = N_C \times N_{\lambda}$, $N_C$ is the number of coronagraph designs, and $N_{\lambda}$ is the number of wavelengths. We mitigate this by parallelizing the exposure time calculation in C using OpenMP. In the end, the exposure time calculation takes 14 seconds for $N=15$ on a 32 core server, a minority of the total run time of the yield code.

\section{Results \& Discussion\label{results_section}}

\subsection{Optimizing the detection bandpass}

We focus first on optimizing the wavelength for exoplanet detection observations. We impose no requirements for spectral characterization until Section \ref{characterization_optimization_section}.

\subsubsection{Scenario A: No wavelength optimization\label{assumptions_section}}

To study the effects of wavelength optimization, we must adopt a fiducial mission description. Currently there is no detailed design for HWO. In lieu of this, we adopt the same performance parameters baselined for the LUVOIR-B mission concept study. Specifically, we adopt a segmented, off-axis primary mirror with an inscribed diameter of 6.7 m. To simplify our study, we adopt only a single coronagraph design, the deformable mirror-assisted charge 6 vortex coronagraph (DMVC) used in the LUVOIR-B study. We adopt all of the same performance parameters given in Table 2 of Ref.~\citenum{stark2019}. We ignore the high-throughput scenario, focusing on the scenario with an end-to-end optical throughput of 0.35 in the second detection band. 

Our goal in this work is to investigate the observation optimization process, not estimate absolute yields. As such, we make several changes to the assumptions in Ref.~\citenum{stark2019}. First, since we will initially focus on detection wavelength optimization and build up to characterization wavelength optimization, we will remove all spectral characterization requirements until Section \ref{characterization_optimization_section}. Whereas Ref.~\citenum{stark2019} assumed the UV channel and vis channel could operate in parallel, we adopt only a single channel operating at a time to simplify the wavelength optimization investigation. Because of this, we move the dichroic split between the UV and visible wavelength channels from 500 nm to 380 nm (discussed further in Section \ref{throughput_section}). Ref.~\citenum{stark2019} required a minimum of six visits to each system to budget for orbit measurement---here we place no requirements on numbers of visits per star to allow the code to fully optimize for completeness. Finally, because the variation of individual stars' exozodi levels may impact wavelength optimization in ways that are difficult to interpret, we opt to assign every star three zodis of dust instead of drawing them randomly from the best fit exozodi distribution from the Large Binocular Telescope Interferometer (LBTI) HOSTS survey\cite{ertel2020}. As a result of these changes, one should not expect the absolute yield numbers to be the same as the LUVOIR-B baseline design.

\begin{figure}[H]
\centering
\includegraphics[width=6in]{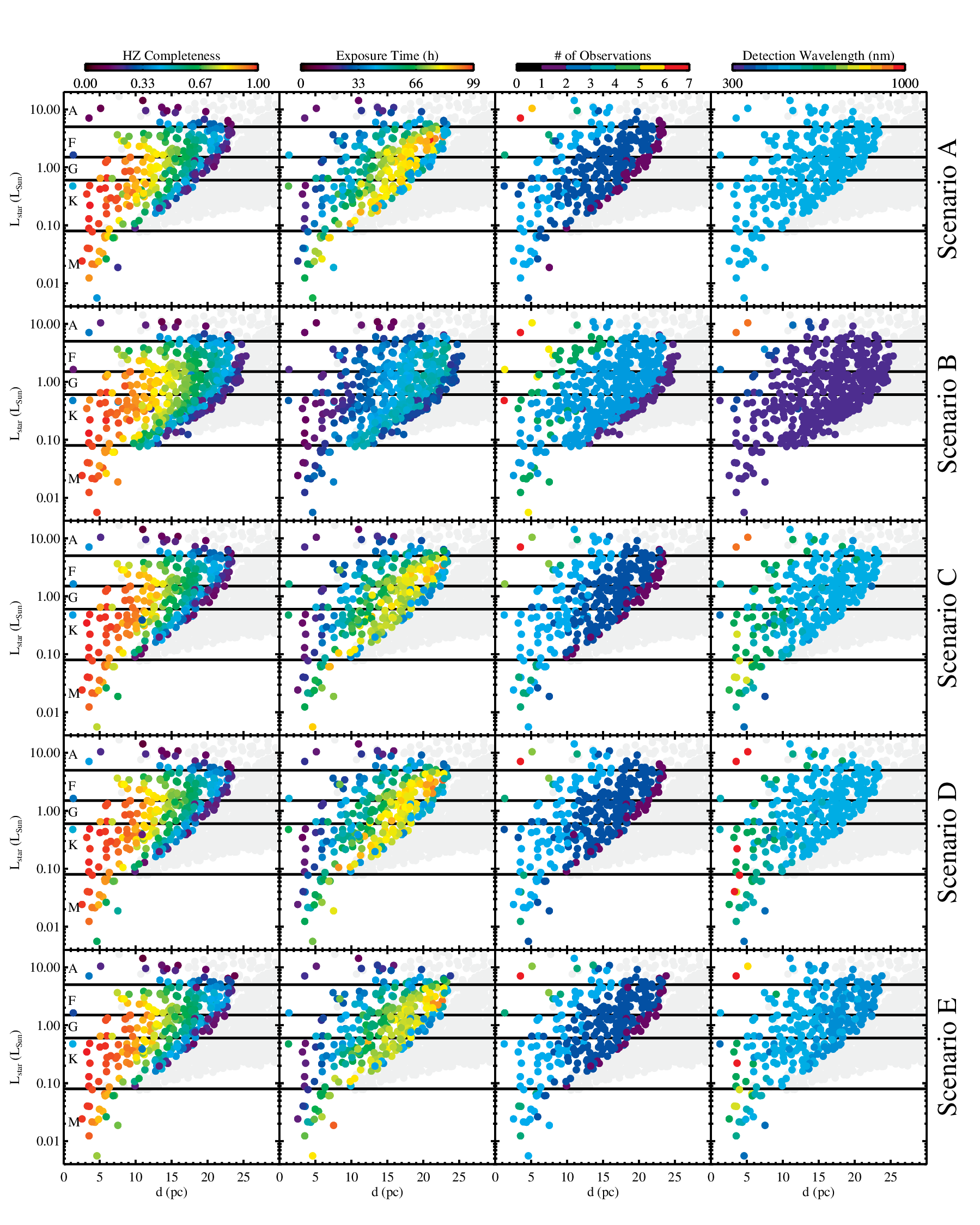}
\caption{Targets selected by our yield code for each of the detection wavelength optimization scenarios considered. From left to right, plots show habitable zone detection completeness, total detection exposure time, number of observations, and detection wavelength chosen for each target. Scenario D is our preferred detection wavelength optimization method; distant targets are preferentially observed in V band while nearby targets are observed at slightly longer wavelengths.  \label{fig:detection_optimization}}
\end{figure}

We calculated the exoplanet yield, optimized for exoEarth candidates (EECs) with no wavelength optimization as a baseline for comparison, which we refer to as Scenario A. Similar to previous works, here EEC yield is calculated as the cumulative completeness of EECs multiplied by $\eta_{\Earth}$, such that results track the expectation value of the yield. We adopted the same definition of an EEC as in Ref.~\citenum{stark2019} and the same value of $\eta_{\Earth} = 0.24$. We assumed a detection wavelength of 500 nm and no spectral characterization requirements. Results are shown in the top row of Figure \ref{fig:detection_optimization}. From left to right, the columns show the targets selected by our yield code color-coded by habitable zone detection completeness, total detection exposure time, number of observations, and detection wavelength chosen for each target. As expected for no wavelength optimization, all targets in the right-most plot are light blue, corresponding to an observation wavelength of 500 nm. The yield of EECs for this baseline scenario is 41.3.

\subsubsection{Scenario B: Wavelength optimization accounting for PSF scale only}

\begin{figure}[H]
\centering
\includegraphics[width=4.5in]{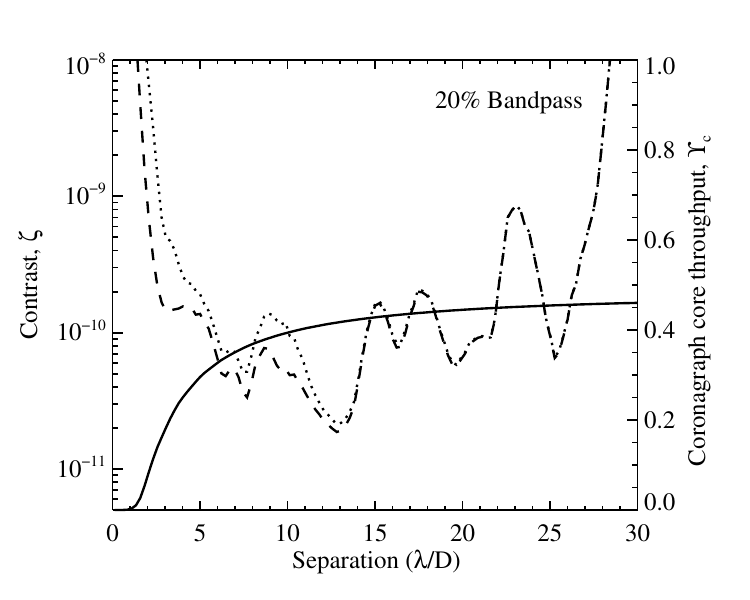}
\caption{Azimuthally-averaged raw contrast $\zeta$ as a function of separation for an on-axis point source (dashed) and an on-axis source with diameter $0.1\lambda/D$ (dotted) for the DMVC charge 6 with an off-axis segmented primary mirror. During yield calculations, we set the contrast to the greater of $\zeta$ and $10^{-10}$.  Adopted core throughput of the planet is shown as a solid line.  \label{DMVC_fig}}
\end{figure}

We expect the scale of the PSF to motivate a shorter detection wavelength. Figure \ref{DMVC_fig} shows the azimuthally averaged performance of the DMVC coronagraph. The contrast of a point source and a typical star with diameter $0.1\lambda/D$ are shown as dashed and dotted lines, respectively. The core throughput of the planet is shown as a solid line. The IWA is formally defined as the separation at which the throughput drops to half its maximum value. In this case, the IWA of the coronagraph is $\sim3.5\lambda/D$. Both contrast and throughput scale with separation in units of $\lambda/D$, the scale of the PSF. As a result, reducing $\lambda$ reduces the on-sky/physical IWA, providing access to more compact HZs.

Decreasing wavelength also reduces the contribution of noise sources. The contributions of the dominant sources of astrophysical noise---leaked starlight, exozodiacal dust, and zodiacal light---all scale linearly with the solid angle of the PSF. This $\lambda^2$ dependence should also push observations to shorter wavelengths.

We repeated our yield calculations from the previous scenario, this time with detection wavelength optimization enabled. We adopted 15 possible detection bandpasses centered on wavelengths equally spaced from 300 - 1000 nm, all with 20\% bandwidth. Here we investigate the impact of PSF scale only. To isolate this effect, we artificially maintained wavelength-independent astrophysical photon arrival rates, a constant optical throughput with wavelength, and a constant planetary reflectance with wavelength. These assumptions will be relaxed one by one in subsequent sections. 

For these calculations we assumed that the diffraction limit of the telescope, as well as the detector pixel scale, adjusted with the wavelength such that the same number of pixels were used for PSF extraction at all wavelengths. The impact of this assumption is negligible, as the adopted detector noise from Ref. \citenum{stark2019} is below levels that would impact broadband detection exposure times\cite{stark2015}. Specifically, a $\sim$6 m telescope receives $\sim$30 photons per minute from an Earth-twin at 10 pc across a 20\% bandpass at visible wavelengths. Assuming even a low 1\% end-to-end throughput implies a signal count rate $> 10^{-3}$ counts pix$^{-1}$ s$^{-1}$, while the dark current of the LUVOIR- and HabEx-baselined EMCCD is $\sim10^{-5}$ counts pix$^{-1}$ s$^{-1}$.

The second row from the top of Figure \ref{fig:detection_optimization} shows the results of enabling wavelength optimization for Scenario B. As expected, nearly all stars prefer the shortest possible wavelength, which reduces astrophysical noise. Notably, two very nearby, early-type stars prefer longer wavelengths. Other than Alpha Cen A, these two stars (HIP 37279 and HIP 97649) have the two largest HZs on the sky (HZ outer edges at 1264 and 1050 mas, respectively). At 500 nm, the $\sim30$ $\lambda/D$ effective OWA of the coronagraph would only probe to distances of 460 mas, providing poor coverage of these stars' HZs. Wavelength optimization is therefore choosing longer wavelengths to access the spatial extent of these stars' HZs. We note that although the HZ of Alpha Cen A is even larger, with an outer edge at 1611 mas, it has a bright nearby companion star, Alpha Cen B, which contributes stray light into the yield calculator's exposure time equation; Alpha Cen A is observed at shorter wavelengths to reduce the stray light from Alpha Cen B.

\subsubsection{Scenario C: Wavelength optimization accounting for PSF scale and stellar spectra\label{section:scenarioC}}

\begin{figure}[H]
\centering
\includegraphics[width=4.5in]{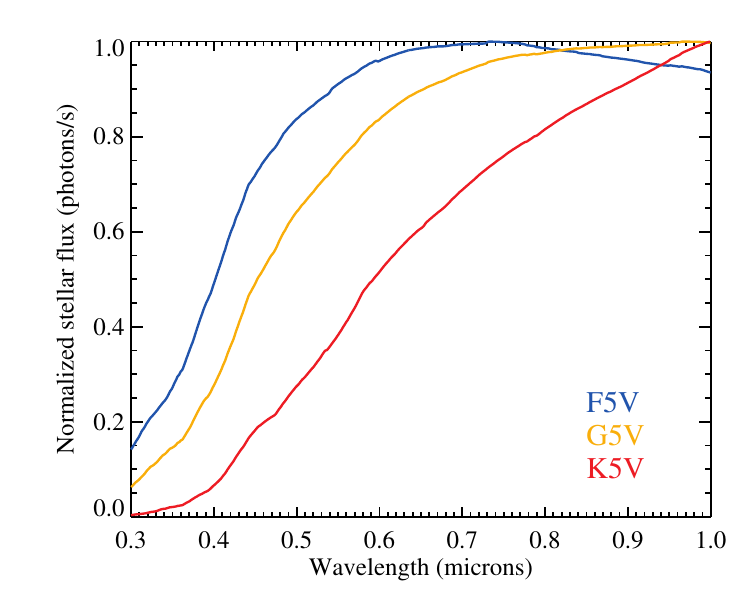}
\caption{The photon arrival rate integrated over a 20\% bandpass as a function of central wavelength. Three spectral types are shown (F5V in blue, G5V in orange, and K5V in red), each normalized to their peak. Stellar spectra will tend to favor longer wavelengths. \label{fig:stellar_spectra}}
\end{figure}

Of course stellar spectra are not constant with wavelength. To illustrate this, we adopted three stellar spectra interpolated from the ATLAS9 models\cite{castelli2003} and integrated the photon arrival rate over a 20\% bandpass as a function of central wavelength. For these three examples, we adopted effective temperatures of 6510, 5660, and 4410 K to represent F5V, G5V, and K5V stars, respectively. For all three, we adopted a turbulence velocity of 2 km s$^{-1}$, a mixing length parameter $l/H = 1.25$, and solar values for the metallicity and $\log g$. Figure \ref{fig:stellar_spectra} shows the normalized bandpass-integrated photon arrival rate for several fiducial spectral types. There is clearly a large penalty for operating near 300 nm, where stars are intrinsically faint. Broadly speaking, the stellar spectra will push the optimized detection wavelength toward 1000 nm.

We repeated our previous yield calculations, but allowed stellar spectra to vary with wavelength by interpolating the photometry in our stellar input catalog (detailed in Ref.~\citenum{stark2019}). The middle row in Figure \ref{fig:detection_optimization} shows the results for Scenario C. Compared to the results in the previous section, stars are now primarily observed at 500 nm. There is a trend in the optimized detection wavelength, with nearby stars preferring slightly longer wavelengths. The EEC yield for this scenario is 42.8, a negligible 1\% increase over our baseline scenario.

We note that while the optimized detection wavelength has a clear trend, it appears somewhat ``noisy." Specifically, six stars within 6 pc have optimized detection wavelengths of 800 nm whereas neighboring stars are optimized at 650 nm. This variation in optimized detection wavelength among neighboring stars in the plot is due to several factors. First, the DMVC exhibits a speckle pattern with several bright, wavelength-dependent spots. We experimented with azimuthally averaging the DMVC contrast map and found that doing so slightly reduces the variation in the optimized detection wavelength plot. However, the dominant factor contributing to this variation appears to be the quality of photometry in our input catalog. As an example, we compare HIP 105090 and HIP 104217. These two stars both have luminosities $\sim0.08$ L$_{\odot}$ and distances of $\sim3.7$ pc, however the former has an optimized detection bandpass of 800 nm while the latter is 650 nm. Manual inspection of the photometry in our input catalog shows that HIP 105090 is $\sim0.6$ mags fainter than HIP 104217 at most bandpasses, except I band, where it is only $\sim0.1$ mag fainter and thus anomalously bright, driving its optimization to 800 nm. Similar issues with I band photometry exist for other low luminosity stars with optimized wavelengths near 1000 nm, highlighting the need for a more accurate input stellar catalog\cite{tuchow2023}. To explicitly test for the impact of the photometry in the stellar input catalog, we temporarily assigned every star the spectrum of a solar twin, calibrated to its apparent V band magnitude. This removed all of the variation from the optimized detection wavelengths. We conclude that the apparent variation in optimized detection wavelength is not a numerical artifact. Rather, it is due to either real astrophysical variations of stellar spectra or imprecise photometry in our input catalog.

\subsubsection{Scenario D: Wavelength optimization accounting for PSF scale, stellar spectra, and optical throughput\label{throughput_section}}

So far we have assumed that the instrument throughput is independent of wavelength. Here we retire that assumption. Ref.~\citenum{stark2019} estimated the end-to-end optical throughput as a function of wavelength for plausible optical layouts assuming a dichroic split between UV and visible wavelength channels at 500 nm. We reproduce that calculation, but shift the dichroic edge from 500 nm to 380 nm, where the UV and vis channel throughputs are equal. The intent of this change is to avoid inserting a sharp transition in the throughput at 500 nm due to outdated assumptions about what wavelengths would be ideal for detection.

\begin{figure}[H]
\centering
\includegraphics[width=4.5in]{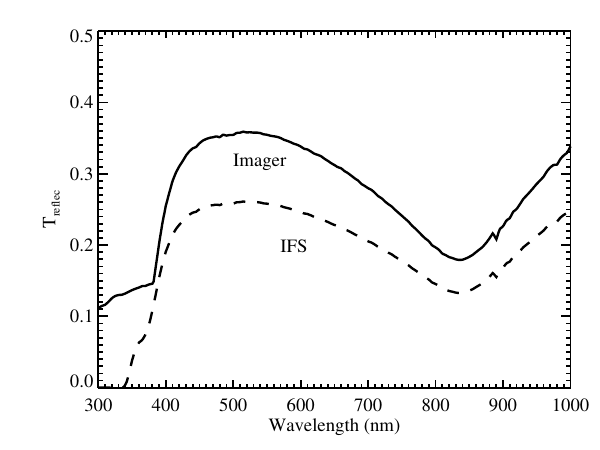}
\caption{The end-to-end reflectivity of all optics as a function of wavelength adopted in this study. Reflectivity for detection and spectral characterization observations are shown as solid and dashed lines, respectively. \label{fig:instrument_throughput}}
\end{figure}

The solid line in Figure \ref{fig:instrument_throughput} shows the instrument throughput as a function of wavelength that we adopt for the imaging mode. To simplify our wavelength optimization investigation, we will treat the UV and vis channels as a single, continuous channel. Figure \ref{fig:instrument_throughput} shows a large penalty when operating short of 400 nm, as well as from 700-900 nm. Instrument throughput therefore favors detection near 500 nm.

\begin{figure}[H]
\centering
\includegraphics[width=4.5in]{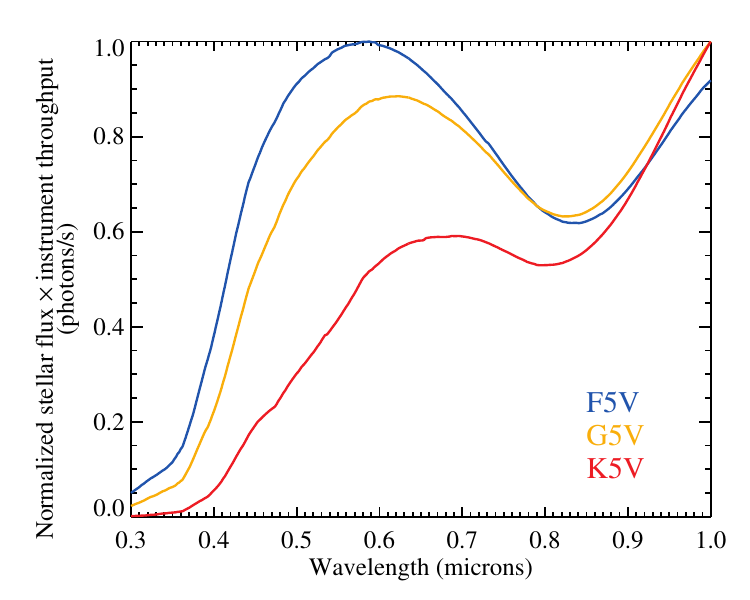}
\caption{The product of the stellar photon arrival rate and instrument throughput curves integrated over a 20\% bandpass as a function of central wavelength, each normalized to their peak. There is a strong spectral preference for observation near V band. \label{fig:stellar_x_instrument}}
\end{figure}

Figure \ref{fig:stellar_x_instrument} shows the combined effects of instrument throughput and stellar spectra. Most of these curves exhibit a local maximum near V band, though it is not necessarily the global maximum. Wavelength optimization methods in yield calculations must therefore be mindful of such local maxima.

We note that the local minimum in the throughput near 800 nm is due to the coated aluminum mirrors prior to the coronagraph required to enable UV science, not the silver optics within the visible channel. The LUVOIR-B design assumed a total of seven aluminum reflections prior to the vis coronagraph channel. Reducing this to two aluminum reflections, which would still enable UV astrophysics, could have a significant impact on these curves, boosting the throughput near 800 nm and shifting the peak of the curves to longer wavelengths.

We repeated our yield calculations from the previous section, but incorporated the wavelength dependence of throughput shown in Fig \ref{fig:instrument_throughput}. The second-to-bottom row of Fig \ref{fig:detection_optimization} shows the results for Scenario D. Most stars that were observed near 800 nm in the previous scenario, where the instrument throughput is low, are now observed at alternative wavelengths. We note that three stars have optimized detection wavelengths of 1000 nm, while neighboring stars have optimized wavelengths of 650 nm. This is due to the imprecise I band photometry of our input catalog discussed in Section \ref{section:scenarioC}, which affects interpolation between I and J bands. These results demonstrate that the wavelength-dependence of the instrument throughput can have a measurable impact on individual observations. However, an EEC yield of 42.6 for this scenario indicates no measurable change in the mission-long science productivity.

This scenario represents our preferred method for detection wavelength optimization, as it incorporates all wavelength-dependent effects without any prior assumptions about the planet's spectral dependence. Next we will consider an additional wavelength optimization effect in which we adopt a wavelength-dependent exoplanet spectrum, though we don't advise implementing this in routine yield calculations as it tunes observations to focus on Earth twins.

\subsubsection{Scenario E: Additional wavelength optimization accounting for planetary reflectance}

\begin{figure}[H]
\centering
\includegraphics[width=6in]{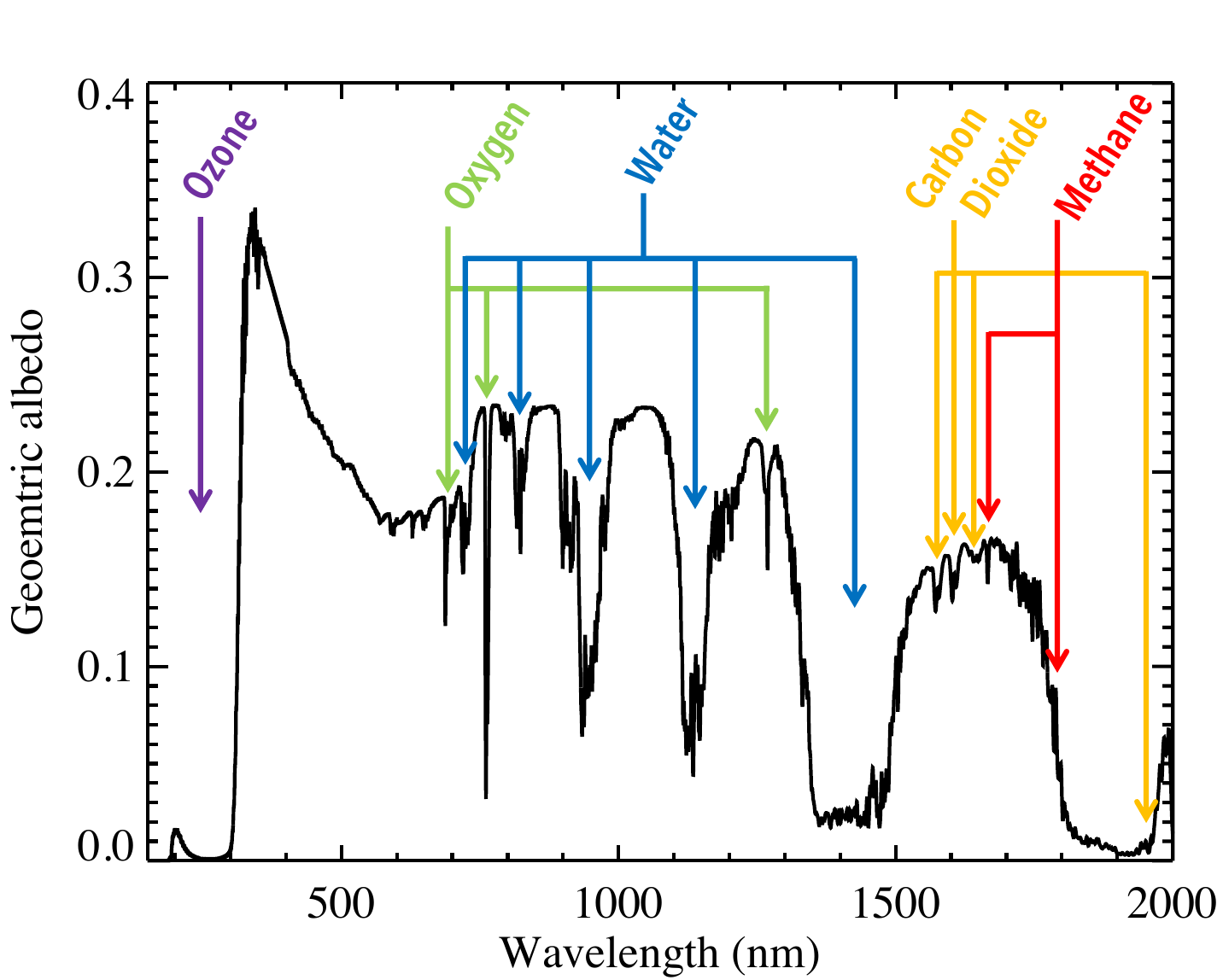}
\caption{The geometric albedo of the Earth with major spectral features labeled\cite{robinson2011,roberge2017}. \label{fig:earth_spectrum}}
\end{figure}

So far we have assumed a wavelength-independent planetary geometric albedo of 0.2. The geometric albedo of Earth, shown in Figure \ref{fig:earth_spectrum}, is $\sim$0.2 at 500 nm\cite{robinson2011,roberge2017}. Rayleigh scattering due to water vapor increases this to $\sim$0.3 at $\sim$350 nm. 

\begin{figure}[H]
\centering
\includegraphics[width=4.5in]{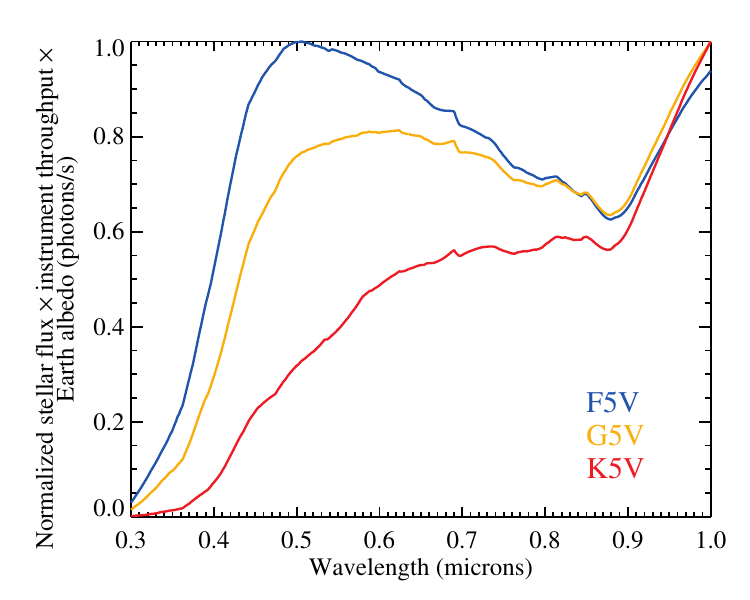}
\caption{The product of the stellar photon arrival rate, instrument throughput, and Earth's geometric albedo integrated over a 20\% bandpass as a function of central wavelength, each normalized to their peak. Compared to Figure \ref{fig:stellar_x_instrument}, the photon arrival rate is enhanced near $\sim$400 nm. \label{fig:stellar_x_instrument_x_planet}}
\end{figure}

If one wanted to optimize the search for potentially habitable planets to true Earth twins, it may be beneficial to take advantage of this Rayleigh scattering peak. Figure \ref{fig:stellar_x_instrument_x_planet} shows the combined effects of the wavelength dependent instrument throughput, Earth's geometric albedo, and stellar spectra. 

To investigate the impact of this effect, we repeated our previous yield calculations from the previous section, but included a wavelength dependent geometric albedo equal to that of the Earth. The bottom row of Figure \ref{fig:detection_optimization} shows the results for Scenario E. The detection wavelength for distant stars switches from 500 nm to 450 nm, and a handful of nearby stars avoid detections at 1000 nm due to the water absorption band. We note that due to the imprecise I band photometry in our input catalog, discussed in Section \ref{section:scenarioC}, a handful of low-luminosity, nearby stars maintain optimized bandpasses longward of 800 nm. The results are otherwise nearly identical to the previous scenario and there is no discernible change in yield. We conclude that, for the instrument parameters assumed here, there is no measurable benefit to including the wavelength dependence of an Earth twin's albedo as a prior when conducting a future survey for Earth-like exoplanets. Table \ref{table:EEC_yields} summarizes the scenarios we have examined.

\begin{deluxetable}{llllll}
\tablewidth{0pt}
\footnotesize
\tablecaption{Summary of scenarios examined\label{table:EEC_yields}}
\tablehead{
\colhead{Scenario} & \colhead{Architecture} & \colhead{Spec.} & \colhead{Sources of wavelength} & \colhead{Wavelength} & \colhead{EEC} \\
\colhead{} & \colhead{} & \colhead{Char.} & \colhead{dependence included} & \colhead{Optimization} & \colhead{Yield} \\
}
\startdata
A & LUVOIR-B & None & None & None & 41.3\\
B &  LUVOIR-B & None & A + PSF scale only & Detection & 65.0\\
C &  LUVOIR-B & None & B + stellar spectra & Detection & 42.8\\
D &  LUVOIR-B & None & C + optical throughput\tablenotemark{*} & Detection & 42.6\\
E &  LUVOIR-B & None & D + planet reflectance & Detection & 42.9\\
F &  LUVOIR-B & H$_2$O & All\tablenotemark{*} & None & 27.9\\
G &  LUVOIR-B & H$_2$O  & All\tablenotemark{*} & Full & 28.2\\
H &  LUVOIR-A & H$_2$O  & All\tablenotemark{*} & None & 56.6\\
I &  LUVOIR-A & H$_2$O  & All\tablenotemark{*} & Full & 58.0\\
J &  LUVOIR-B\tablenotemark{**} & H$_2$O  & All\tablenotemark{*} & None & 4.8\\
K &  LUVOIR-B\tablenotemark{**} & H$_2$O & All\tablenotemark{*} & Full & 10.7\\
\enddata
\vspace{-0.1in}
\tablenotetext{*}{Scenario D, ignoring planet reflectance, defines ``all" wavelength variation.}
\tablenotetext{**}{Adopted the wavelength-dependent QE of the Roman CGI EMCCD.}
\end{deluxetable}

\subsection{Optimizing the characterization bandpass\label{characterization_optimization_section}}

So far we have only considered exoplanet detections. Here we require some minimum spectral characterization observation for every detected planet. We calculate probabilistic spectral characterization times in the same manner as Ref. \citenum{stark2015}. We remind the reader that, as discussed in Section \ref{assumptions_section}, we place no minimum requirement on the number of observations per star. While this formally departs from the LUVOIR and HabEx operations concepts, for which six visits were required to determine the planet's orbit prior to spectral characterization\cite{luvoirfinalreport,habexfinalreport}, we continue this assumption here, as previous studies have shown this requirement to have a minor impact on the yield of coronagraph-based missions \cite{stark2016} and we focus on the impact of wavelength optimization methods, not absolute yields.

\label{assumptions_section}

\subsubsection{Scenario F: No wavelength optimization}

We first consider a baseline scenario that is similar to the assumptions made in the LUVOIR study\cite{luvoirfinalreport}. For each detected EEC, we require a spectral characterization observation to detect water vapor. We adopt a spectral resolution of $R=140$ and S/N$=5$ evaluated at the continuum at a wavelength of 1000 nm. This is a bit more conservative than the LUVOIR study, which adopted $R=70$\cite{luvoirfinalreport}. We adopt the same throughput and number of spectral characterization pixels as Ref. \citenum{luvoirfinalreport}. For this baseline scenario, we turn off all wavelength optimization, including detection wavelength optimization. The top row of Figure \ref{fig:characterization_optimization} shows the results. As expected, all stars are detected at 500 nm and characterized at 1000 nm. The EEC yield for this baseline scenario is 27.9.

\subsubsection{Scenario G: Full wavelength optimization\label{section:full_opt}}

The LUVOIR study adopted the detection of water at 1000 nm as its baseline characterization requirement for all EECs. The water absorption feature at 950 nm is relatively deep, but it is not the only water absorption feature available. As shown in Fig. \ref{fig:earth_spectrum}, there are multiple water lines at shorter wavelengths. Although these lines are not as prominent, their shorter wavelength makes them potentially useful for targets that are limited by the IWA of the coronagraph. By providing the yield code multiple options for the detection of water, we may be able to expand the target list and increase yields.

To determine the required S/N necessary to detect H$_2$O as a function of wavelength, we performed spectral retrieval analyses on modern Earth-twins. We used the Bayesian retrieval application PSGnest\footnote{https://psg.gsfc.nasa.gov/apps/psgnest.php}, which is part of the Planetary Spectrum Generator (PSG) \cite{PSG,PSGbook}. PSGnest uses a grid-based Bayesian nested sampling algorithm derived from MultiNest, with pre-built spectral grids produced using the PSG radiative transfer model to efficiently and quickly run thousands of retrievals; for a full description of the spectra calculation, grid-building, and analysis, see Ref. \citenum{susemiehl23}. We generated our fiducial spectra using the modern Earth-like values used in Ref. \citenum{feng18} and an isotropic cloud model with 50\% clear and 50\% cloudy spectra. We adopted a spectral resolution of $R$=140, a 20\% bandpass, and considered bandpasses centered every 17 nm from 661--900 nm. We established detection strength using the log-Bayes Factor, where log-Bayes $> 5$ is considered a strong detection\cite{benneke13}. We then varied the S/N from 3 -- 16 in steps of one, and determined the minimum S/N that produced a strong detection of H$_2$O. This way, we thoroughly investigate the change in feature detectability and the resultant strength at various wavelengths and S/N. For a description of the full S/N analysis and further abundance studies, see Ref. \citenum{latouf23}.

\begin{figure}[H]
\centering
\includegraphics[width=4.5in]{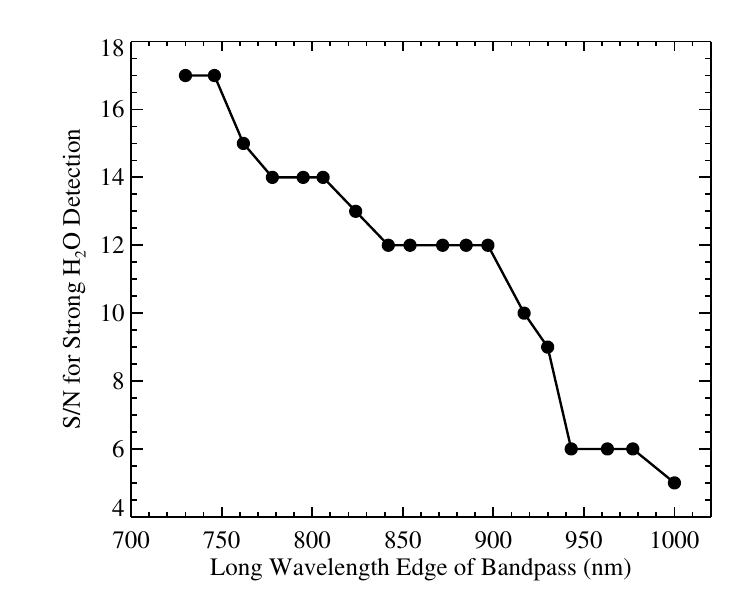}
\caption{The S/N required for a strong water vapor detection on an Earth twin as a function of long-wavelength edge for a 20\% bandpass and spectral resolution $R=140$. \label{fig:lambda_vs_snr}}
\end{figure}

Figure \ref{fig:lambda_vs_snr} shows the results of our spectral retrieval analysis. To ensure that the planet is visible over the entire bandpass, we adopt the long-wavelength edge of the bandpass instead of the bandpass center. H$_{2}$O is most easily detectable at wavelengths in excess of 900 nm. At shorter wavelengths, H$_{2}$O is still strongly detectable, but requires higher S/N. 

We independently verified these results using Markov chain Monte Carlo (MCMC) sampling at two points in the wavelength grid, one representing a long wavelength example at 1000 nm and one representing a short wavelength example at 780 nm. We used rfast to infer an atmospheric model via radiative transfer forward modeling, incorporating a simplified instrument noise model and a Bayesian MCMC statistical analysis tool \cite{RobinsonSalvador2023}. We adopted the same spectral resolution and bandpass coverage for each of the two H$_2$O detection wavelengths and simulated directly imaged views of an Earth-like exoplanet at S/N of 5 and 10 for the 1000 nm and 780 nm grid points, respectively. We assumed constant noise estimates specified at the shortest wavelength in each bandpass. Our atmospheric inputs for generating a fiducial spectrum with the rfast forward model differ slightly from the previous grid experiments. Instead of adapting the constant atmospheric profiles from Ref. \citenum{feng2018}, we generated self-consistent, steady-state atmospheric profiles of a modern Earth-sun twin using the photochemical component of the Atmos model \cite{arney2016,arney2017}. This distinction is important, as the abundance of H$_2$O is known to vary with altitude.

\begin{figure}[H]
\centering
\includegraphics[width=4.5in]{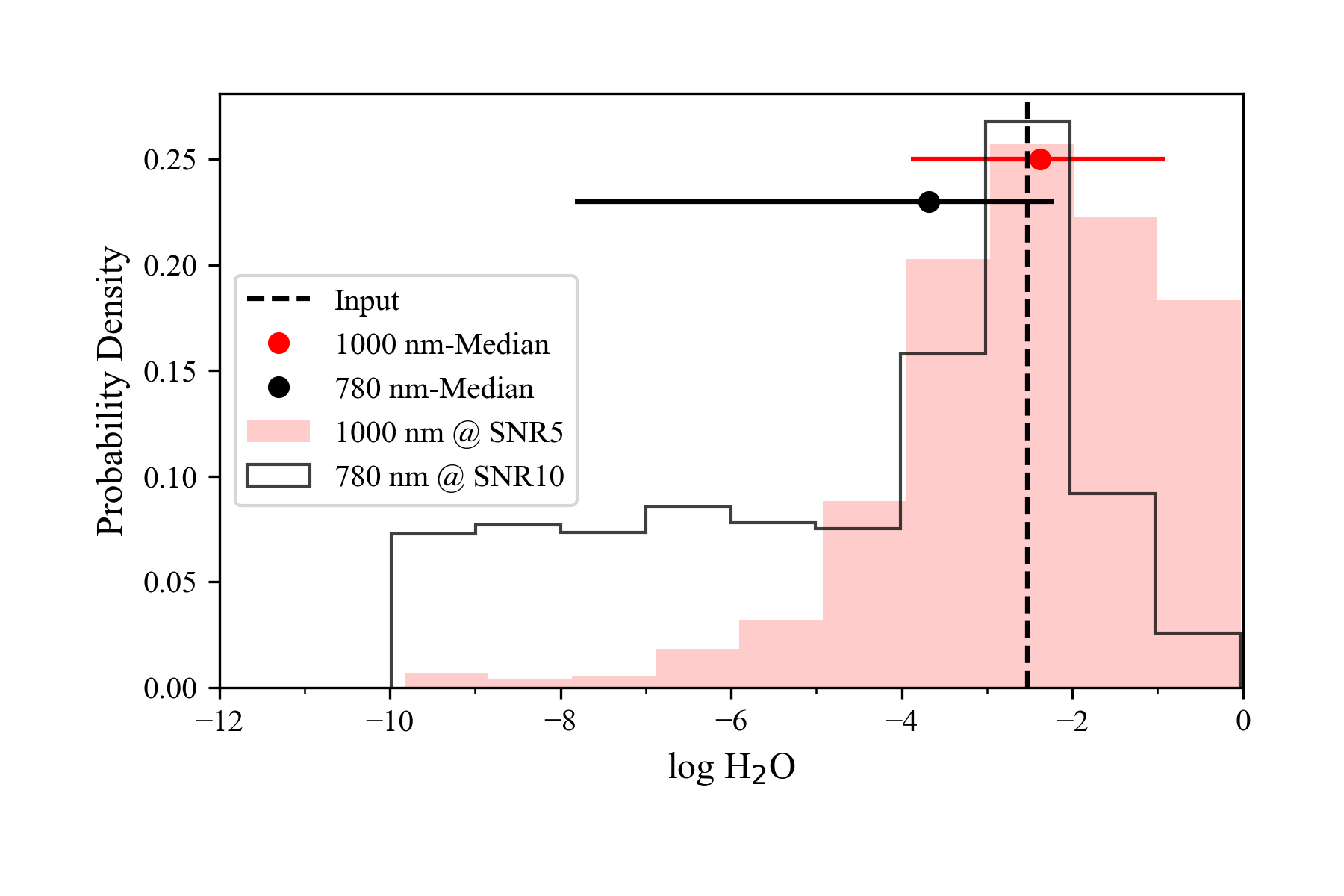}
\caption{Marginal posterior distribution for H$_2$O at the two wavelengths chosen for verification. A strong detection is made at both wavelengths. \label{fig:h2o_verification}}
\end{figure}

In Figure \ref{fig:h2o_verification}, we show the marginal posterior distribution for H$_2$O at the two wavelengths chosen for verification. Both distributions exhibit clear detections, which are peaked near the original input abundance (a volume mixing ratio of 3e-3), but also have statistically significant tails extending to lower abundance values. Also shown are the 1-$\sigma$ confidence intervals taken from the cumulative distribution of each posterior. These retrieval results are consistent with the aforementioned grid analysis when taking into account independent analyses using two different retrieval models with differing altitude abundance dependencies.

Finally, we are ready to consider ``full wavelength optimization." Here we enable both detection wavelength optimization and characterization wavelength optimization. Detection wavelength optimization is implemented in the same fashion as Scenario D above. For characterization wavelength optimization, we provide the code with the wavelengths, spectral resolutions, and S/N required to detect water shown in Figure \ref{fig:lambda_vs_snr}. Like the LUVOIR study, we assume that the IFS lenslet array is Nyquist sampled at 500 nm and estimate the number of detector pixels as a function of wavelength as $N_{\rm det,pix} = N_{\rm lenslets} N_{\rm dispersed}$, assuming $ N_{\rm lenslets} = 4 \left(\lambda_{\rm char}/500\; {\rm nm}\right)^2$ and $N_{\rm dispersed}=6 \left(140/R\right)$, where we account for six detector pixels per lenslet per spectral resolution element at the native detector spectral resolution of 140.

\begin{figure}[H]
\centering
\includegraphics[width=6.5in]{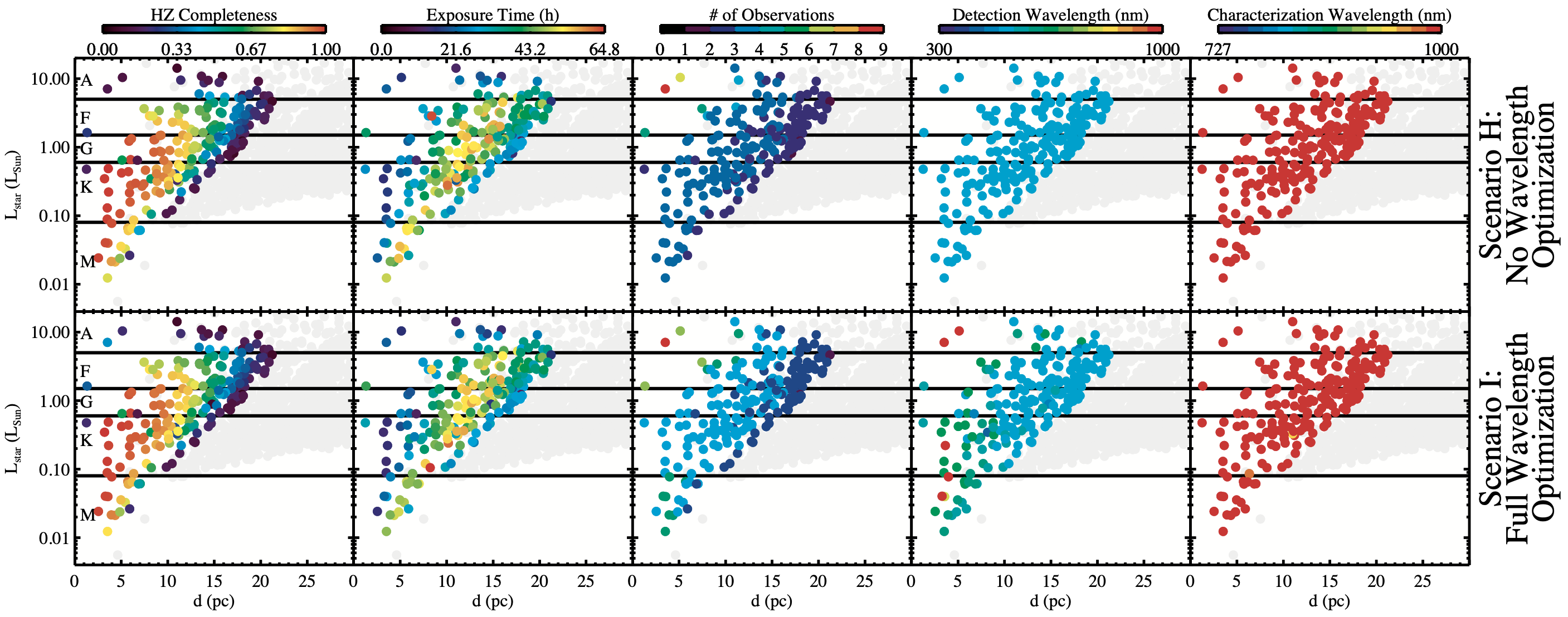}
\caption{LUVOIR-B targets selected by our yield code for the scenario with no wavelength optimization (top row) and full wavelength optimization (bottom row). From left to right, plots show habitable zone detection completeness, total detection exposure time, number of observations, detection wavelength chosen for each target, and characterization wavelength chosen for each target. We confirm that the detection and characterization wavelength assumptions made in the LUVOIR-B study were largely correct.\label{fig:characterization_optimization}}
\end{figure}

The bottom row of Figure \ref{fig:characterization_optimization} shows the results for Scenario G. In spite of providing the yield code with multiple bandpass options for the detection of water, the 1000 nm bandpass was selected for nearly every star and the EEC yield remains effectively unchanged. The reason for this is that the increase in required S/N at shorter wavelengths is too costly to make these bands useful. More precisely, reducing the wavelength for spectral characterization from 1000 nm to 780 nm roughly triples the required S/N, which would increase exposure time by a factor of nine, all else being equal. For the shorter wavelength band to be useful, the benefits of noise reduction (due to a smaller PSF) and the throughput gain (by operating at a larger working angle in units of $\lambda/D$) must exceed this factor of four. For the DMVC, which has a very graceful throughput curve near the IWA (as shown in Fig. \ref{DMVC_fig}), a factor of 0.78 change in working angle never amounts to a factor of nine in throughput. Our results show that for LUVOIR-B, the detection and characterization wavelength assumptions made in the LUVOIR study were largely correct.

\subsection{The utility of wavelength optimization\label{section:utility}}

So far our results have confirmed the assumptions made for the LUVOIR-B study and have not shown significant benefit from wavelength optimization. However, this is likely not the case in general. The details of the instrument may play a significant role in determining the utility of wavelength optimization. Here we discuss several scenarios in which wavelength optimization can benefit the mission.

\subsubsection{Scenarios H \& I: LUVOIR-A}

\begin{figure}[H]
\centering
\includegraphics[width=4.5in]{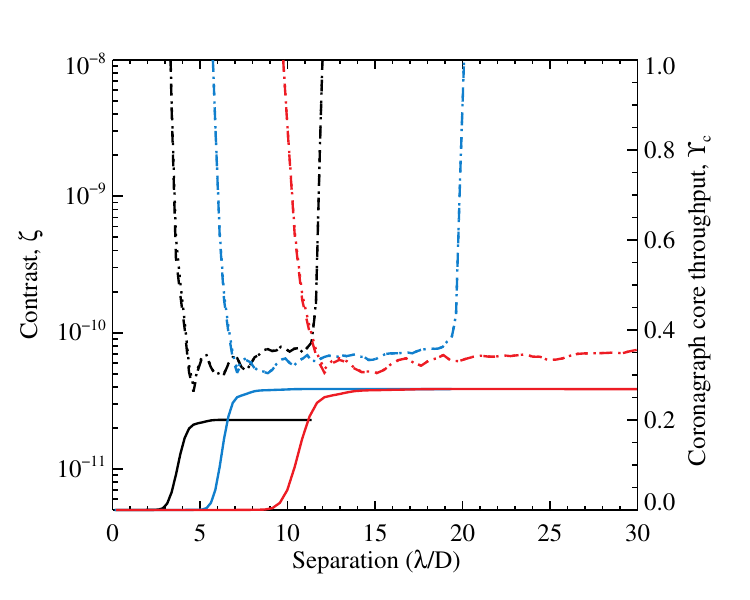}
\caption{Azimuthally-averaged raw contrast $\zeta$ as a function of separation for an on-axis point source (dashed) and an on-axis source with diameter $0.1\lambda/D$ (dotted) for the three APLC designs adopted for LUVOIR-A. During yield calculations, we set the contrast to the greater of $\zeta$ and $10^{-10}$.  Adopted core throughput of the planet is shown as a solid line.  \label{APLC_fig}}
\end{figure}

First, as previously discussed, the gentle slope of the DMVC throughput curve near the IWA effectively precludes the utility of shorter wavelength water bands for characterization. Not all coronagraphs have such gentle IWAs. Figure \ref{APLC_fig} shows the throughputs and contrasts for the three Apodized Pupil Lyot Coronagraphs (APLCs) adopted for the LUVOIR-A architecture\cite{stlaurent2018}. The throughput curves for these coronagraphs resemble step functions; operating interior to the IWA is not an option like it is for the DMVC. To study the utility of wavelength optimization for an APLC, we performed two simulations of the 15 m LUVOIR-A, with and without wavelength optimization, referred to as Scenarios H and I, respectively. We considered the three APLCs, but did not include the Apodized Vortex Coronagraph used in the LUVOIR-A study\cite{luvoirfinalreport}. We made assumptions similar to Ref. \citenum{luvoirfinalreport}, but adopted a single coronagraph channel at a time and modified the dichroic split between channels as described in Section \ref{assumptions_section}. 

\begin{figure}[H]
\centering
\includegraphics[width=6.5in]{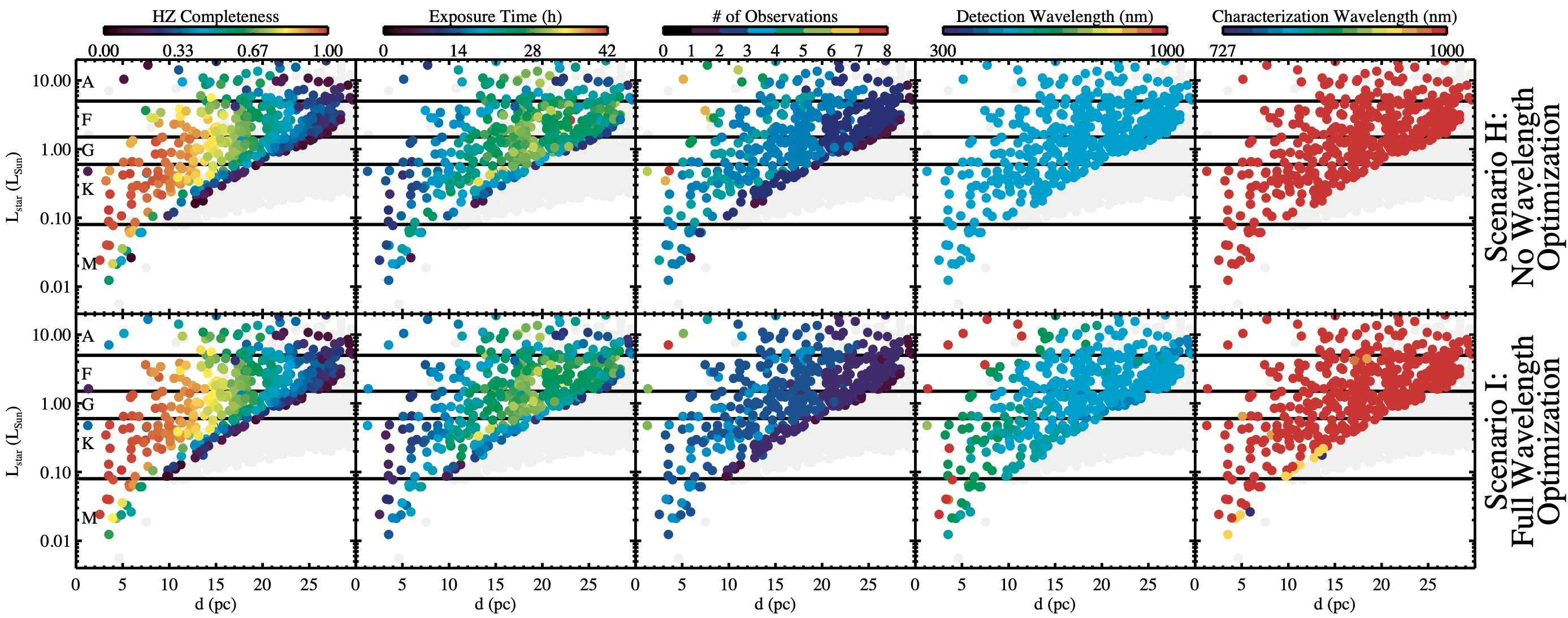}
\caption{LUVOIR-A targets selected by our yield code for the scenario with no wavelength optimization (top row) and full wavelength optimization (bottom row). Plotting conventions are the same as Fig \ref{fig:characterization_optimization}. Wavelength optimization allows the mission to search for water at shorter wavelengths for planets around K type stars that would be inaccessible at 1000 nm.\label{fig:luvoira}}
\end{figure}

Figure \ref{fig:luvoira} shows our results with wavelength optimization applied to LUVOIR-A with APLC coronagraphs. A number of K stars with HZs near, or interior to, the coronagraph IWA at 1000 nm are instead characterized at shorter wavelengths. While this case demonstrates the utility of wavelength optimization, the overall impact on yield is a modest 3\% increase.

\subsubsection{Scenarios J \& K: LUVOIR-B with wavelength-dependent QE}

The wavelength dependence of the end-to-end throughput can also play a major role in wavelength optimization. As shown in Section \ref{throughput_section}, reductions in instrument throughput at certain wavelengths can shift the optimal detection bandpasses. The LUVOIR study adopted a wavelength-independent detector quantum efficiency (QE) of 90\% from 500 -- 1000 nm. The \emph{Roman} Coronagraph Instrument's (CGI) EMCCD QE drops dramatically at wavelengths longer than 800 nm, to just $6.8\%$ at 980 nm. 

\begin{figure}[H]
\centering
\includegraphics[width=4.5in]{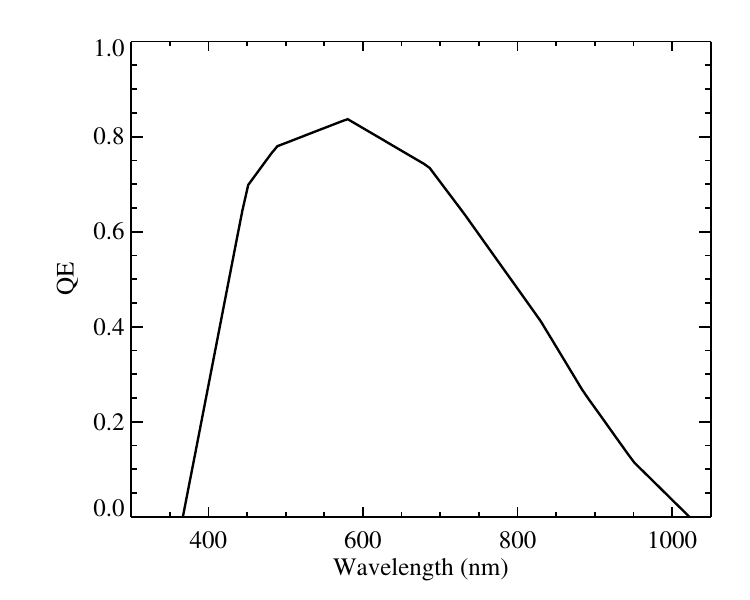}
\caption{Linearly-interpolated QE from the Roman CGI requirements. \label{fig:qe}}
\end{figure}

\begin{figure}[H]
\centering
\includegraphics[width=6.5in]{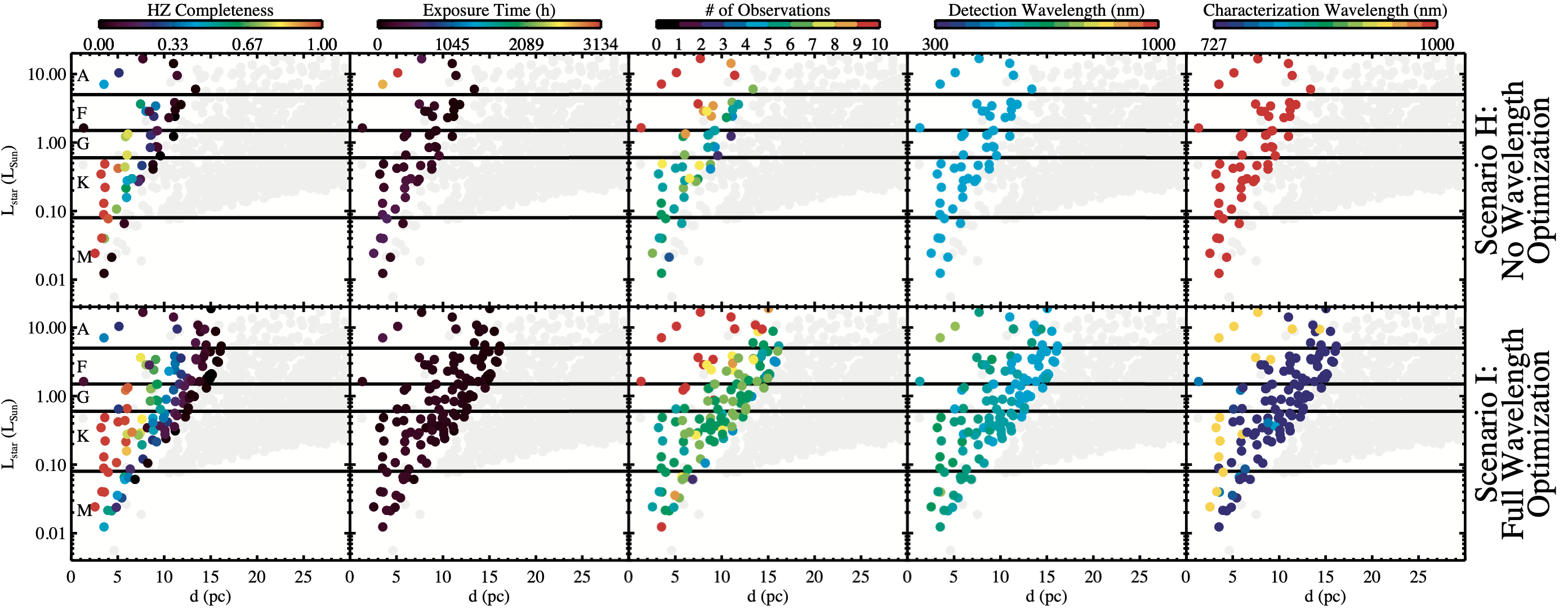}
\caption{LUVOIR-B targets selected by our yield code for the scenario with no wavelength optimization (top row) and full wavelength optimization (bottom row) when adopting the Roman CGI detector QE shown in Figure \ref{fig:qe}. Plotting conventions are the same as Fig \ref{fig:characterization_optimization}. The QE curve shifts most characterization and detection wavelengths well short of 1000 nm.\label{fig:luvoirb_qe}}
\end{figure}

To examine the impacts of such a QE on wavelength optimization, we adopted the QE curve shown in Figure \ref{fig:qe}, which was linearly interpolated from the Roman CGI EMCCD QE requirements\footnote{\url{https://roman.ipac.caltech.edu/sims/Param_db.html}}. With this new QE curve, we ran two additional yield calculations for LUVOIR-B with and without wavelength optimization, referred to as Scenarios J and K, respectively. As shown in Figure \ref{fig:luvoirb_qe}, the low QE near 1000 nm results in all characterization wavelengths shifting to shorter values. Here wavelength optimization improves the yield by a factor of 2.2 relative to no optimization. Of course the low yield in the case of no wavelength optimization is due to forcing spectral characterizations at 1000 nm. In reality, if a mission concept had such a low QE at 1000 nm one would not force spectral characterization at this wavelength, and would instead alter the characterization bandpass. However, this would require us to intuit the ideal bandpass for spectral characterization for the entire population of stars---the wavelength optimization methods presented here automate this decision for us.

The utility of characterization wavelength optimization depends on the interplay between instrument performance, the target list, and the combinations of wavelength, S/N, and $R$ required to detect a given atmospheric absorption feature. In this paper we limited our analysis to the spectral characterization of a single molecular species: water. However, these same techniques can be applied to additional species like oxygen and carbon dioxide, which will exhibit different relative trades. Future spectral retrieval studies should provide the combinations of wavelength, S/N, and $R$ required for all of these species such that they can be incorporated into yield studies. Future yield studies may also be able to optimize sequential spectral characterization observations, allowing us to determine the ``full" spectral characterization times on all targets to detect a broad range of molecules, as well as test observational strategies to search for life like that laid out in Figure 1-5 of the LUVOIR Final Report\cite{luvoirfinalreport}.

\section{Conclusions}

We implemented detection and characterization wavelength optimization methods in the AYO yield code. We used these methods to explicitly confirm the observational wavelength assumptions made for the LUVOIR-B study, namely that the optimum detection wavelength is 500 nm for most stars and the optimum wavelength to detect water is near 1000 nm, given LUVOIR-B's assumed instrument performance parameters. We showed that including the wavelength dependent albedo of an Earth-twin as a prior provides no significant benefit to the yields of exoEarth candidates, supporting the assumption of a wavelength-independent albedo for EEC yield calculations. We also demonstrated how wavelength-dependent instrument performance can impact the optimum wavelengths for detection and characterization. The optimization methods we established automate wavelength selection for the mission, helping to adapt observations to the performance parameters of future exoplanet imaging mission concepts. We expect these methods to play an important role for trade studies for the Habitable Worlds Observatory.

\label{conclusions}

\section{Data and Code Availability}
NASA regulations govern the release of source code, including what can be released and how it is made available. Readers should contact the corresponding author if they would like copies of the visualization software or data produced for this study.

\acknowledgments

This project was supported by the NASA HQ-directed ExoSpec work package under the Internal Scientist Funding Model (ISFM). N.~L. gratefully acknowledges support from an NSF GRFP. The authors would like to thank two anonymous referees for feedback that improved this manuscript.

%\bibliography{bibliography.bib}
%\bibliographystyle{spiejour}   % makes bibtex use spiejour.bst
\bibliography{ms_v5.bbl}

\listoffigures
\listoftables

\end{spacing}
\end{document}